\documentclass[11pt]{article}
\usepackage{a4}
\usepackage[dvips]{epsfig}
\usepackage{latexsym}
\usepackage{graphicx}
\usepackage{psfrag}
\def\setb@se#1{\baselineskip=#1 \normalbaselineskip=#1}
\setb@se{14pt}
\def\setb@se#1{\baselineskip=#1 \normalbaselineskip=#1}
\setb@se{14pt}

\newcommand{\be}{\begin{equation}}
\newcommand{\ee}{\end{equation}}
\newcommand{\beqn}{\begin{eqnarray}}
\newcommand{\eeqn}{\end{eqnarray}}
\newcommand{\bsub}{\begin{subeqnarray}}
\newcommand{\esub}{\end{subeqnarray}}
\newcommand{\disp}{\displaystyle}

\newcounter{subequation}[equation]

\makeatletter \expandafter\let\expandafter\reset@font\csname
reset@font\endcsname
\newenvironment{subeqnarray}
  {\arraycolsep1pt
    \def\@eqnnum\stepcounter##1{\stepcounter{subequation}{\reset@font\rm
      (\theequation\alph{subequation})}}\eqnarray}%
  {\endeqnarray\stepcounter{equation}}
\makeatother

\begin{document}

\title{ {\bfseries On the number of instabilities of cosmological solutions
in an Einstein-Yang-Mills system.}}
\author{P\'eter Forg\'acs, S\'{e}bastien Reuillon
\\Laboratoire de Math\'{e}matiques et Physique Th\'{e}orique
\\CNRS-UMR 6083\\Universit\'{e} de Tours, Parc de Grandmont\\37200 Tours, France}

\date{}
\maketitle

\begin{abstract}

A detailed numerical stability analysis of the static, spherically
symmetric globally regular solutions of the Einstein-Yang-Mills
equations with a positive cosmological constant, $\Lambda$, is
carried out.
It is found that the number of unstable modes in the even
parity sector is $n$ for solutions with $n=1$,$2$
nodes as $\Lambda$ varies.
The solution with $n=3$ nodes exhibits a rather surprising behaviour in that
the number of its unstable modes {\sl jumps} from
$3$ to $1$ as $\Lambda$ crosses (from below) a critical value.
In particular the topologically $3$-sphere type solution with $n=3$ nodes
has only a {\sl single} unstable mode.
\end{abstract}

\newpage

As we have learned during the last few years, the coupling of
nonabelian gauge fields to Einstein's gravity leads to a rather
large class of solutions of potential physical interest,
such as ``gravitating glueballs'' of Bartnik-McKinnon \cite{bm}, and
``hairy black holes'' \cite{kunzle,voga,biz}.
These {\sl globally regular} solutions are static, spherically symmetric,
and they form a discrete family labelled by the number of nodes, $n$, of the
(single and purely ``magnetic'') Yang-Mills amplitude.
Furthermore the above solutions have been generalized to a continuous family
by allowing for a {\sl positive} cosmological constant, $\Lambda$ \cite{cosmo}.
These ``cosmological'' solutions exist only for a limited range of the cosmological constant,
$0<\Lambda\leq\Lambda_{\scriptstyle\rm reg}(n)$, $n$ being the node
number of the Yang-Mills amplitude.

The important problem of stability has been studied mostly in linear perturbation
theory, with the conclusion that all of these solutions are unstable
\cite{instab,lav-mai,volkov-instab1,volkov-instab2,mavro}.
The spectrum of small fluctuations can be divided with respect to parity into
even (``gravitational'') and odd (``sphaleronic'') sectors. Within the most general
spherically symmetric ansatz, mostly numerical results give convincing evidence
that a gravitating glueball or black hole with $n$ nodes possesses exactly $n$
unstable modes in the even parity sector \cite{instab,lav-mai}, and in Refs.\
\cite{volkov-instab1,volkov-instab2}
analytic arguments have been presented that there are also precisely
$n$ unstable modes in the odd parity sector.

\indent The aim of this paper is to point out that
in the even parity sector the number of unstable modes of the cosmological solutions
shows an unexpected dependence on the cosmological constant, $\Lambda$,
contrary to a previous claim \cite{volkov-instab2}.
Since the geometry of the cosmological solutions changes substantially for a certain
value of $\Lambda=\Lambda_{\rm crit}(n)$, as then an equator develops,
i.e.\ the surface area, $4\pi r^2$,
of the invariant 2-spheres becomes stationary for a certain value of $r$ and then it
starts to decrease.
A careful numerical analysis reveals that the number of unstable modes in the even
parity sector is $1$ resp.\ $2$ for solutions with $n=1$ resp.\ $n=2$
nodes independently of $\Lambda$ as it varies from $0$ to $\Lambda_{\scriptstyle \rm reg}(n)$.
The number of unstable modes of the $n=3$ nodes solution exhibits, however,
a rather surprising behaviour. As $\Lambda$ tends to a certain value,
$\Lambda_{\scriptstyle \rm merg}\approx0.2658$,
two eigenvalues of the unstable modes approach each other, actually merging
at $\Lambda_{\scriptstyle \rm merg}$ and then they simply {\sl disappear} from the
spectrum.
In particular the globally regular, spatially compact
and topologically $3$-sphere-type solution (where $\Lambda=\Lambda_{\rm reg}(n)$)
with $n=3$ zeros has only a {\sl single} unstable mode.
This is a somewhat unexpected phenomenon, since the
linearized perturbation equations can be transformed to a standard $1$ dimensional
Schr\"odinger eigenvalue problem for the perturbation, $\delta W$, of the Yang-Mills
background for a particular choice of the coordinates (Schwarzschild gauge).
The unstable modes are then nothing but bound states of the corresponding Schr\"odinger
equation.
Now for such a Schr\"odinger problem one does not expect that the values of
bound-state
energy levels could even coincide as a single parameter ($\Lambda$)
is varied, let alone {\sl disappear}
from the spectrum without crossing the zero energy threshold.
The Schwarzschild gauge is, however, singular
when an equator is present, which is always the case for
$\Lambda\in[\Lambda_{\scriptstyle \rm crit}(n),\Lambda_{\scriptstyle\rm reg}(n)]$.
In that case the potential in the associated
Schr\"odinger equation becomes so singular that the standard theory of self adjoint
operators is no longer applicable.
It is quite remarkable that the number of unstable modes can actually change
as the cosmological constant varies and
it would be clearly desirable to achieve a better understanding of this phenomenon.

The most general spherically symmetric line element can be written as
\be
\label{metric}
ds^2=e^{2\nu(R,t)}dt^2-e^{2\lambda(R,t)}dR^2-r^2(R,t)d\Omega^2\,,
\ee
and the minimal spherically symmetric Ansatz for the YM field
is given by
\be\label{YMansatz}
 A=\left(T^1W(R,t)d\theta+T^2
W(R,t)\sin\theta d\phi+T^3\cos\theta d\phi\right)\,,
\ee
where the $T^i$ $(i=1,2,3)$ denote an orthonormal set of Lie algebra generators of SU(2).
For the above Ans\"atze the (suitably rescaled) Einstein equations take the form:
\begin{subeqnarray}\label{rteqs}
1-U-2V-e^{-2\lambda}{\nu}'(r^{2})'
+e^{-\nu}({(r^2)_{\scriptscriptstyle ,t}}e^{-\nu})_{, \scriptscriptstyle t}&=&0\,,\\
1+U-2V-e^{-\lambda}(e^{-\lambda}(r^2)')'
+e^{-2\nu}{\lambda_{,\scriptscriptstyle t}}{(r^2)_{,\scriptscriptstyle t}}
&=&0\,,\\
e^{-\nu}({r_{, \scriptscriptstyle t}}e^{-\nu})_{, \scriptscriptstyle t}
-e^{-\lambda}(r'e^{-\lambda})'-e^{-\nu-\lambda}[(e^{\nu-\lambda}r\nu')'
-(e^{\lambda-\nu}r{\lambda_{, \scriptscriptstyle t}})_{, \scriptscriptstyle t}]
-\frac{\partial V}{\partial r}&=&0\,,\\
{r'_{,\scriptscriptstyle t}}+r'{\lambda_{,\scriptscriptstyle t}}
 -{r_{, \scriptscriptstyle t}}\nu'+\frac{2{W_{,\scriptscriptstyle t}}W'}{r}&=&0\,,
 \end{subeqnarray}
 together with the Yang-Mills field equation for the amplitude $W$:
\be\label{ymeq}
(e^{\nu-\lambda}W')'-(e^{\lambda-\nu}{W_{,\scriptscriptstyle t}})_{,\scriptscriptstyle t}
-e^{\nu+\lambda}\frac{W(W^{2}-1)}{r^{2}}=0\,,\ \ee
where
$$
V=\frac{(1-W^{2})^{2}}{2r^{2}}+\frac{\Lambda r^{2}}{2}\,,
\quad U=e^{-2\nu}({r_{,\scriptscriptstyle t}}^2-2 {W_{, \scriptscriptstyle t}}^{2})
+e^{-2\lambda}(r'^{2}-2W'^{2})\,,
$$
and $r_{,\scriptscriptstyle t}:= {\partial r \over\partial t}$,
while prime denotes ${\partial\over\partial R}$.

Following Ref.\ \cite{bfm1} the static field eqs.\ are conveniently written as
\bsub\label{stateqs}
\dot{r}&=&N,\\
\dot{N}&=&(\kappa-N)\frac{N}{r}-2\frac{\dot{W}^{2}}{r}\,,\\
\dot{\kappa}&=&\left[1-\kappa^{2}+2\dot{W}^{2}-2\Lambda r^2\right]/r\,,\\
\ddot{W}&=&W\frac{W^2-1}{r^{2}}-(\kappa-N)\frac{\dot{W}}{r}\,,\
\esub
where from now on $\dot{f}:={d f\over d\sigma}:= e^{-\lambda}f'$, and
$N$, $\kappa$ are {\sl defined} as $N:=\dot r$, $\kappa:=N+r\dot{\nu}$.
The ($RR$) component of the Einstein eqs.\ (\ref{rteqs}a) yields
the constraint :
\be\label{constr} 1+N^2+2\dot{W}^2-2V=2\kappa N\,. \ee
Solutions of Eqs.\ (\ref{stateqs}) with a regular origin have the following
power series expansion
\be\label{bco}
\begin{array}{lll}
r=\sigma-\frac{1}{3}(2b^2+\frac{\Lambda}{6})\sigma^3+O(\sigma^5)\,,&
\kappa=1+(2b^2-\frac{\Lambda}{2})\sigma^2+O(\sigma^4)\,,
&\nu=(2b^2-\frac{\Lambda}{6})\sigma+O(\sigma^3)\,,\\
& \\
N=1-(2b^2+\frac{\Lambda}{6})\sigma^2+O(\sigma^4)\,, & W=1-b\sigma^2+O(\sigma^4)\,,\\
\end{array}
\ee
where $b$ is a free parameter. We consider solutions with
a cosmological horizon, characterized in our notations
by the vanishing of $N(\sigma_h)$ and a simple pole of $\kappa(\sigma)$ at
$\sigma=\sigma_h$.
The corresponding series expansions of the functions near the cosmological horizon
are given as:
\be\label{bch}
\begin{array}{lll}
r=r_h+r_h^{(2)}x^2+O(x^4)\;, & \disp\kappa=\frac{-r_h}{x}+\kappa_h^{(1)}x+O(x^3)\,,
&\nu=\nu_h+\ln(x)+O(x^2)\,,\\
 & \\
N=N_h^{(1)}x+O(x^3)\;, & W=W_h+W_h^{(2)}x^2+O(x^4)\,, \\
\end{array}
\ee
with $x=\sigma_h-\sigma$, and
\be\label{expr}
\begin{array}{ll}
r_h^{(2)}=-\disp\frac{N_h^{(1)}}{2}\;, & \kappa_h^{(1)}=\disp\frac{1}{3}\left(2\Lambda r_h-
\frac{1}{r_h}+\frac{N_h^{(1)}}{2}\right)\,, \\
 & \\
N_h^{(1)}=\disp\frac{1}{2r_h}\left(\frac{(1-W_h^2)^2}{r_h^2}+\Lambda r_h^2-1\right)\;, &
W_h^{(2)}=-\disp\frac{1}{4}W_h\frac{1-W_h^2}{r_h^2}\,, \\
\end{array}
\ee
where $W_h$, $\nu_h$ and $r_h$ are free parameters.

As found in Ref.\ \cite{cosmo} the cosmological solutions naturally fall
into three qualitatively different classes as the cosmological constant varies
from $\Lambda=0$ to $\Lambda=\Lambda_{\rm reg}(n)$.
All of these solutions may be indexed by the number of zeros, $n$, of the
Yang-Mills amplitude, $W$.
For $\Lambda=0$ the space-time
corresponding to the globally regular Bartnik-McKinnon solutions
is asymptotically flat. As soon as $\Lambda>0$
the globally regular cosmological solutions
correspond to asymptotically de Sitter space-times
(with a cosmological horizon),
as long as the cosmological constant does not exceed
a critical value, i.e.\ $0<\Lambda<\Lambda_{\rm crit}(n)$, constituting the first class.
The geometry of the solutions changes when $\Lambda$ exceeds
$\Lambda_{\rm crit}(n)$, since
then {\sl an equator} develops outside of the cosmological horizon, i.e.\
the radius of the invariant 2-spheres, $r(\sigma)$, reaches a maximum and then it
decreases down to zero where a geometrical singularity develops.  The equator
is located outside of the cosmological horizon for  $\Lambda<\Lambda_*(n)$.
As $\Lambda$ further increases,
$\Lambda_*(n)<\Lambda<\Lambda_{\rm reg}(n)$, the position of the equator moves
inside the horizon. Such solutions with an equator correspond to a
singular geometry of the ``bag of gold'' type.
Finally for a special value $\Lambda=\Lambda_{\rm reg}(n)$ the position
of the horizon coincides with the singularity and then remarkably, there exists
a {\sl globally regular} solution corresponding to a space-time whose spatial
sections are {\sl compact}. No regular solution seems to exist outside
this range of the cosmological constant. For $n=1$, this
solution is analytically known \cite{Cervero-Hosotani}:
\be\label{analsol}
 \kappa=N=W=\cos(\sigma/\sqrt 2)\,,\quad r=\sqrt 2
\sin(\sigma/\sqrt 2)\,,\quad \Lambda_{\scriptstyle \rm reg}(1)=3/4,
\ee
corresponding to a static Einstein universe whose spatial section are {\sl 3-spheres},
with a constant energy density due to the Yang-Mills fields.

To study the linear stability problem of the above solutions, one
introduces time dependent perturbations of the form
\be
f(\sigma,t)=f(\sigma)+\delta f(\sigma)e^{i\omega t}\,,
\ee
where now $f$ stands for any of the functions $\{r,N,\kappa,W\}$,
and one linearizes the field eqs.\ (\ref{rteqs}a-d) around a static
background, $f(\sigma)$.
In this way one obtains the following set of linear differential eqs.\
for the perturbations:
\begin{subeqnarray}\label{lineqs}
\dot{\delta r}&=&\frac{\kappa-N}{r}\delta r-\frac{2\dot{W}}{r}\delta W+N\delta\lambda\,,\\
\dot{\delta\kappa}&=&(\dot{\kappa}-e^{-2\nu}\omega^2r)\delta\lambda
-2\frac{\kappa-N}{r}\delta\kappa+4\frac{\ddot{W}}{r}\delta
W-\left(4\Lambda+\frac{3\dot{\kappa}}{r}+e^{-2\nu}\omega^2\right)\delta r\,,\\
\ddot{\delta W}&=&-\frac{\kappa-N}{r}\dot{\delta
W}+\left(\frac{3W^2-1}{r^2}
-\frac{2\dot{W}^2}{r^2}-e^{-2\nu}\omega^2\right)\delta
W-2\frac{\ddot{W}}{r}\delta
r-\frac{\dot{W}}{r}\delta\kappa\nonumber\\
&
&+\left(2\ddot{W}+\frac{\kappa-N}{r}\dot{W}\right)\delta\lambda+\dot{W}\dot{\delta\lambda}\,,
\
\end{subeqnarray}
with the linearized constraint (\ref{rteqs}a):
\begin{equation}\label{constrlin}
N\delta\kappa-(\dot{\kappa}-e^{-2\nu}\omega^2r)\delta
r+2\ddot{W}\delta W-2\dot{W}\dot{\delta W}
+2\dot{W}^2\delta\lambda=0\,.
\end{equation}
We remark that Eq.\ (\ref{lineqs}a) is nothing but
the $(t,R)$ component of the linearized Einstein equations.
The system (\ref{lineqs}) is invariant with respect to the following gauge (diffeomorphism)
transformations:
\be\label{gaugetransf}
\begin{array}{ll}
\delta\lambda\longrightarrow\delta\lambda+\dot{F}\;, &
\delta\kappa\longrightarrow\delta\kappa+(\dot{\kappa}-e^{-2\nu}\omega^2r)F\,, \\
 & \\
\delta r\longrightarrow\delta r +NF\;, & \delta W\longrightarrow\delta W +\dot{W}F\,, \\
\end{array}
\ee
where $F$ is a function of the variable $\sigma$. As already
mentioned, due to the presence of an equator
the stability analysis of the cosmological
solutions is more complicated than that of
asymptotically flat ones.
When there is no equator there exists a particular regular gauge --
the Schwarzschild gauge -- defined by $\delta r=0$,
in which the gravitational degrees of freedom can be eliminated
from the equation of the Yang-Mills perturbation amplitude
in Eq.\ (\ref{lineqs}c).
Indeed, in this gauge one obtains just a standard Schr\"odinger equation:
\be\label{lineqschw}
-e^{\nu}\left(e^{\nu}\dot{\delta
W}\right)^{\displaystyle\bf \cdot}+V_{\scriptscriptstyle \rm S}\delta W=\omega^2\delta W\,,
\ee
where the potential, given by
\be\label{potschw}
V_{\scriptscriptstyle \rm S}=e^{2\nu}\frac{3W^2-1}{r^2}+4e^{\nu}\left(e^{\nu}
\frac{\dot{W}^2}{rN}\right)^{\bf\displaystyle \cdot}\,,
\ee
becomes {\sl singular} when the solution has an equator,
$V_{\scriptscriptstyle \rm S}\propto1/(\sigma-\sigma_{\rm eq})^2$,
due to the vanishing of $N$.
The unstable mode of the analytically known 3-sphere background (\ref{analsol})
corresponds to the following solution of the Schr\"odinger equation (\ref{lineqschw}):
\be \delta\label{sol-potschw}
W=\sin(\sigma/\sqrt{2})\tan(\sigma/\sqrt{2}),\qquad\omega^2=-1.
\ee
Note that the solution (\ref{sol-potschw}) is not even square integrable
because of its divergence at the equator $\sigma_{\rm eq}=\pi/\sqrt 2$. According to
the classical theory of self-adjoint operators one should impose boundary conditions
at the equator which would, however, just exclude the solution (\ref{sol-potschw})
(see Ref.\ \cite{volkov-instab2} for a discussion on this point), so it seems that
the correct interpretation of the singular Schr\"odinger equation (\ref{lineqschw})
in the present context is beyond the standard theory.
It is clear for physical reasons that one {\sl cannot} impose  boundary
conditions for the perturbations at the equator, since the singularity there
is an artifact due to the gauge choice.

On the other hand one can simply choose a {\sl globally regular gauge}
where the equator is a regular point of Eqs.\ (\ref{lineqs}).
The only disadvantage of such regular gauges is that one
does not obtain a Schr\"odinger-type perturbation equation, but a
somewhat unusual eigenvalue problem. It is then not easy to
interpret the eigenfunctions and give a meaning to the number of nodes
as in the case of a Schr\"odinger equation.
Nevertheless we find it both conceptually clearer and also better suited for
our numerical procedure to work in a globally regular gauge, which we have chosen
to be $\delta\lambda=0$.

The solutions of Eqs.~(\ref{lineqs}) corresponding to locally regular perturbations
admit the following power series expansions at the origin:
\be\label{linbco}
\delta r(\sigma)=\frac{4b}{3}\delta W_2\sigma^3+O(\sigma^5)\,, \quad
\delta\kappa(\sigma)=-4b\delta W_2\sigma^2+O(\sigma^4)\,,\quad
\delta W(\sigma)=\delta W_2\sigma^2+O(\sigma^4)\,,
\ee
and near the horizon
\be\label{linbch}
\delta r(x)=\delta r_h x^a(x^2+O(x^4))\,,\quad
\delta\kappa(x)=\delta\kappa_h x^a(x+O(x^3))\,,\quad
\delta W(x)=\delta W_h x^a(1+O(x^2))\,,\
\ee
with the following definitions
\be\label{linexpr}
a^2=-e^{-2\nu_h}\omega^2\,,\quad
\delta r_h=-\frac{4W_h^{(2)}}{r_h(a+1)}\delta W_h\,,\quad
\delta\kappa_h=-\frac{(a-1)^2}{(a+3)}\delta r_h\,,
\ee
where $\delta W_2$, $\omega^2$ and $\delta W_h$ are free
parameters. For the background (\ref{analsol}) the unstable mode
can be written in the $\delta\lambda=0$ gauge as:
\be
\delta W=-\delta\kappa=(\sigma/\sqrt2)
\sin(\sigma/\sqrt 2)\,,\quad \delta r=\sqrt 2\sin(\sigma/\sqrt
2)-\sigma\cos(\sigma/\sqrt 2)\,.
\ee

We have numerically integrated Eqs.~(\ref{lineqs})
by a fifth order adaptive step-size Runge-Kutta method
from both the
origin and the horizon subject to the boundary conditions~(\ref{linbco})
and~(\ref{linbch}) in a given static background solution,
and by varying
the free parameters we have matched the solution at an intermediate point
(shooting to a fitting point).

Our main numerical results are presented on Figures 1~$(n=1,2,3)$ and in Tables
1--4.
The $\Lambda$-dependence of the unstable mode of the $n=1$ background
solution displayed on Fig.\ 1($n=1$) agrees with that of Ref.\ \cite{volkov-instab2}.
The main new behaviour of the unstable modes is exhibited on Fig.\ 1~$(n=3)$:
the largest eigenvalue unstable mode with {\sl two} nodes of the
$n=3$ node background solution
-- corresponding to the highest energy bound state
wave function in the Schr\"odinger description --
{\sl looses} one of its nodes at $\Lambda_{\scriptstyle \rm merg}\simeq 0.2657847$ where
its eigenvalue coincides with that of the {\sl single node} bound state.
At $\Lambda_{\scriptstyle \rm merg}$ the coinciding eigenvalues
are $\omega_2^2=\omega_1^2\approx -69$.
For $\Lambda>\Lambda_{\scriptstyle \rm merg}$ these modes with nodes completely
disappear from the spectrum, so only a single unstable mode
{\sl without nodes} (the lowest lying bound state) survives.
It is worth noting that the lowest lying eigenvalues of the bound states
for the $n=2$ and in particular for the $n=3$ node background solution
exhibit quite large variations in a rather small $\Lambda$ interval close to
$\Lambda=0$.
For example the energy of the lowest lying bound state
in the $n=3$ background drops from $\approx -3000$ to $\approx -190000$
as $\Lambda$ varies from $0$ to $\approx 0.02$. For backgrounds with
$n>3$ this phenomenon is getting more and more pronounced,
making numerical integration quite difficult.

Preliminary investigations \cite{MaiBrei} suggest that in fact for all $n>3$ node solutions
there is such a $\Lambda_{\scriptstyle \rm merg}(n)$
where precisely {\sl two} negative eigenvalues collide, leading to a diminution of the
unstable modes by two. Therefore it is natural to conjecture that the number of
instabilities of the $n\ge 3$ node topologically $3$-sphere-type solution
in the even parity sector is in fact $n-2$.

Finally for completeness we have also computed the eigenvalues of the
unstable modes in the {\sl odd parity} sector,
in a regular gauge for solutions with $\Lambda=0$ and for
the topologically 3-sphere solutions.
Omitting all computational details here, we just state that our results
show that the number of unstable modes
for both $\Lambda=0$ and $\Lambda=\Lambda_{\scriptstyle \rm reg}(n)$
is $n$ (up to $n=3$),
in agreement with the analytical arguments of Ref.\ \cite{volkov-instab2}.
The numerical eigenvalues are provided in Table 4.

\vskip0.2truecm
\noindent
{\bfseries Acknowledgements}
\vskip0.1truecm
We would like to thank P.~Breitenlohner and D.~Maison for useful discussions and
in particular for verifying in great detail our numerical results.
We also thank M.~S.~Volkov for discussions and for checking our results for the 3-sphere.


\begin{table}[h]
\parbox{12cm}{\caption{Numerical values of the parameters for some $n=1$ solutions.}}
 \vskip 0.5cm
\begin{tabular}{|c|c|c|c|c|} \hline
$\Lambda$ & $b$ & $r_h$ & $W_h$ & $\omega_0^2$\\
\hline
0     & 0.45371627 & $\infty$ & -1 & -3.284026\\
0.001 & 0.45358471 & 53.924708 & -0.983515 & -3.283974\\
0.01  & 0.45234499 & 16.431260 & -0.947872 & -3.287678\\
0.1   & 0.43582245 &  4.441700 & -0.859362 & -3.162423\\
0.5   & 0.32415673 &  0.956828 & -0.904637 & -1.649222\\
0.7   & 0.26406135 &  0.208989 & -0.990741 & -1.097542\\
0.75  & 0.25       &     0     &     -1    & -1       \\
\hline
\end{tabular}
\end{table}
\begin{table}[h]
\parbox{12cm}{\caption{Numerical values of the parameters for some $n=2$ solutions.}}
\vskip 0.5cm
\begin{tabular}{|c|c|c|c|c|c|} \hline
$\Lambda$ & $b$ & $r_h$ & $W_h$ & $\omega_1^2$ & $\omega_0^2$\\
\hline
0     & 0.65172553 & $\infty$ & 1 & -17.7898 &-93.9115 \\
0.001 & 0.65357281 & 53.775678 & 0.837449 & -20.0993 & -100.6350\\
0.01  & 0.66188220 & 16.271892 & 0.560250 & -39.3607 & -153.0908\\
0.1   & 0.63599416 &  4.224264 & 0.332182 & -76.9804 & -222.8341\\
0.3   & 0.49283156 &  1.244954 & 0.647999 & -13.2317 &  -17.9967\\
0.35  & 0.44326287 &  0.419923 & 0.938400 &  -5.7734 &   -9.5080\\
0.36424423 & 0.42959769 & 0 & 1 & -4.7046 & -8.0863\\
\hline
\end{tabular}
\end{table}
\begin{table}[h]
\parbox{12cm}{\caption{Numerical values of the parameters for some $n=3$ solutions.}}
\vskip 0.5cm
\begin{tabular}{|c|c|c|c|c|c|c|} \hline
$\Lambda$ & $b$ & $r_h$ & $W_h$ & $\omega_2^2$ & $\omega_1^2$ & $\omega_0^2$\\
\hline
0     & 0.69704005 & $\infty$ & -1 & -53.3132 & -390.5425 & -2946.8374\\
0.001 & 0.70174503 & 53.754661 & -0.321763 & -370.5305 & -1235.9470 & -14402.84\\
0.01  & 0.70073671 & 16.257493 & -0.119067 & -764.4160 & -9097.367 & -94777.96\\
0.1   & 0.66137928 &  4.206730 & -0.075888 & -493.7848 & -9938.45 & -89162.65\\
0.26578472 & 0.54488550 & 1.351120  & -0.444434 & -69.0027 & -69.0678 & -153.7337\\
0.29  & 0.51312783 &  0.347933 & -0.947303 & - & - & -35.7286\\
0.29321764 & 0.50882906 &  0 & -1 & - & - & -30.7787\\
\hline
\end{tabular}
\end{table}
\begin{table}[h!]
\parbox{12cm}{\caption{Eigenvalues of the unstable modes in the odd parity sector.}}
\vskip 0.5cm
\begin{tabular}{|c|c|c|} \hline
$n$ & $\Lambda=0$ & $\Lambda=\Lambda_{\scriptstyle \rm reg}(n)$\\
\hline
1 & -3.872828 & -1\\
\hline
2 & -24.083467 & -3.665216\\
  & -82.480704 & -4.619788\\
\hline
3 & -73.653692 & -7.938386\\
  & -325.34471 & -9.198014\\
  & -3005.4050 & -33.599200\\
\hline
\end{tabular}
\end{table}
\begin{figure}[h]
\vbox{
\hbox to\linewidth{\hss%
    \psfrag{omega2}{$\omega_0^2$}
    \psfrag{Lambda}{$\Lambda$}
    \psfrag{n=1}{$n=1$}
    \resizebox{7.5cm}{6.5cm}{\includegraphics{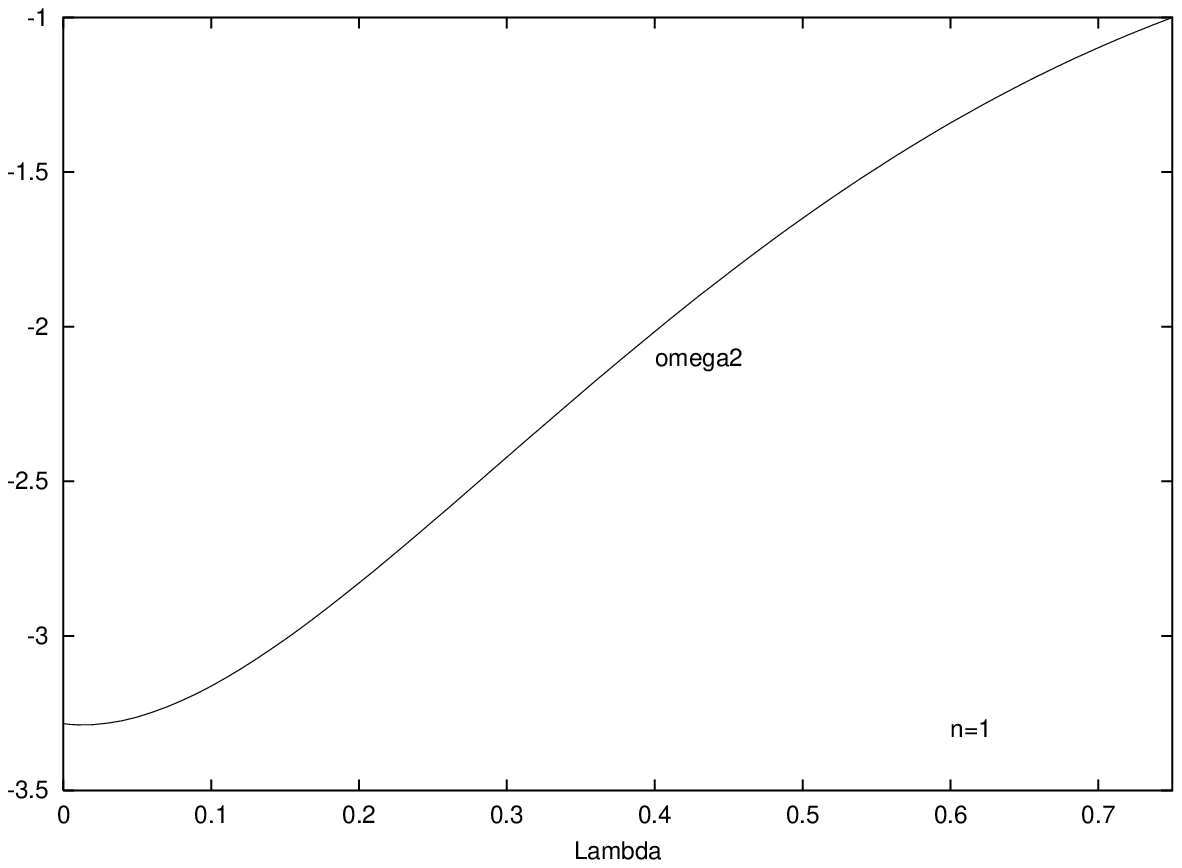}}%
\hspace{5mm}%
    \psfrag{10e-4}{{\scriptsize $10^{-4}$}}
    \psfrag{10e-3}{{\scriptsize $10^{-3}$}}
    \psfrag{10e-2}{{\scriptsize $10^{-2}$}}
    \psfrag{omega1^2}{$\omega_1^2$}
    \psfrag{omega2^2}{$\omega_0^2$}
    \psfrag{Lambda}{$\Lambda$}
    \psfrag{n=2}{$n=2$}
        \resizebox{7.5cm}{6.5cm}{\includegraphics{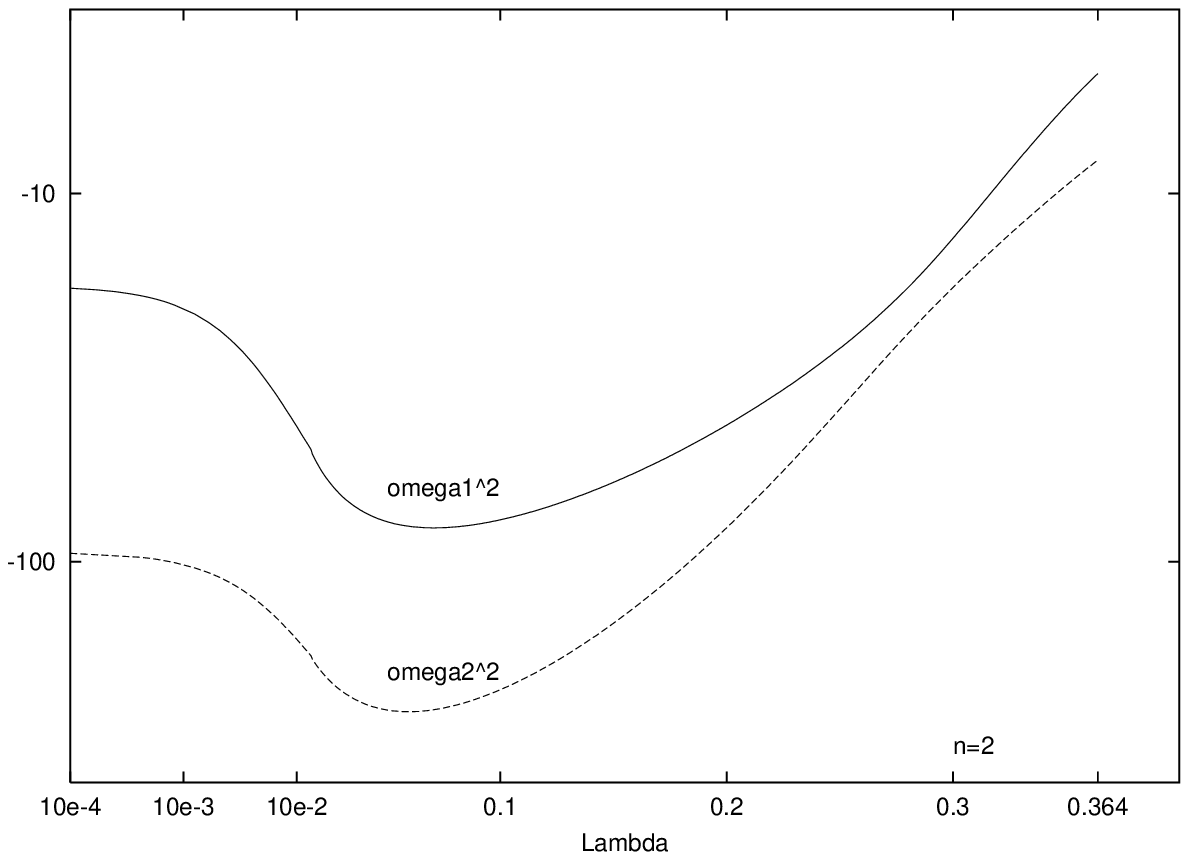}}%
\hss} \vspace{3mm}
\hbox to\linewidth{\hss%
    \psfrag{10e-4}{{\scriptsize $10^{-4}$}}
    \psfrag{10e-3}{{\scriptsize $10^{-3}$}}
    \psfrag{10e-2}{{\scriptsize $10^{-2}$}}
    \psfrag{-10e4}{{\scriptsize $-10^{4}$}}
    \psfrag{-10e5}{{\scriptsize $-10^{5}$}}
    \psfrag{omega1^2}{$\omega_2^2$}
    \psfrag{omega2^2}{$\omega_1^2$}
    \psfrag{omega3^2}{$\omega_0^2$}
    \psfrag{Lambda}{$\Lambda$}
    \psfrag{n=3}{$n=3$}
    \resizebox{7.5cm}{6.5cm}{\includegraphics{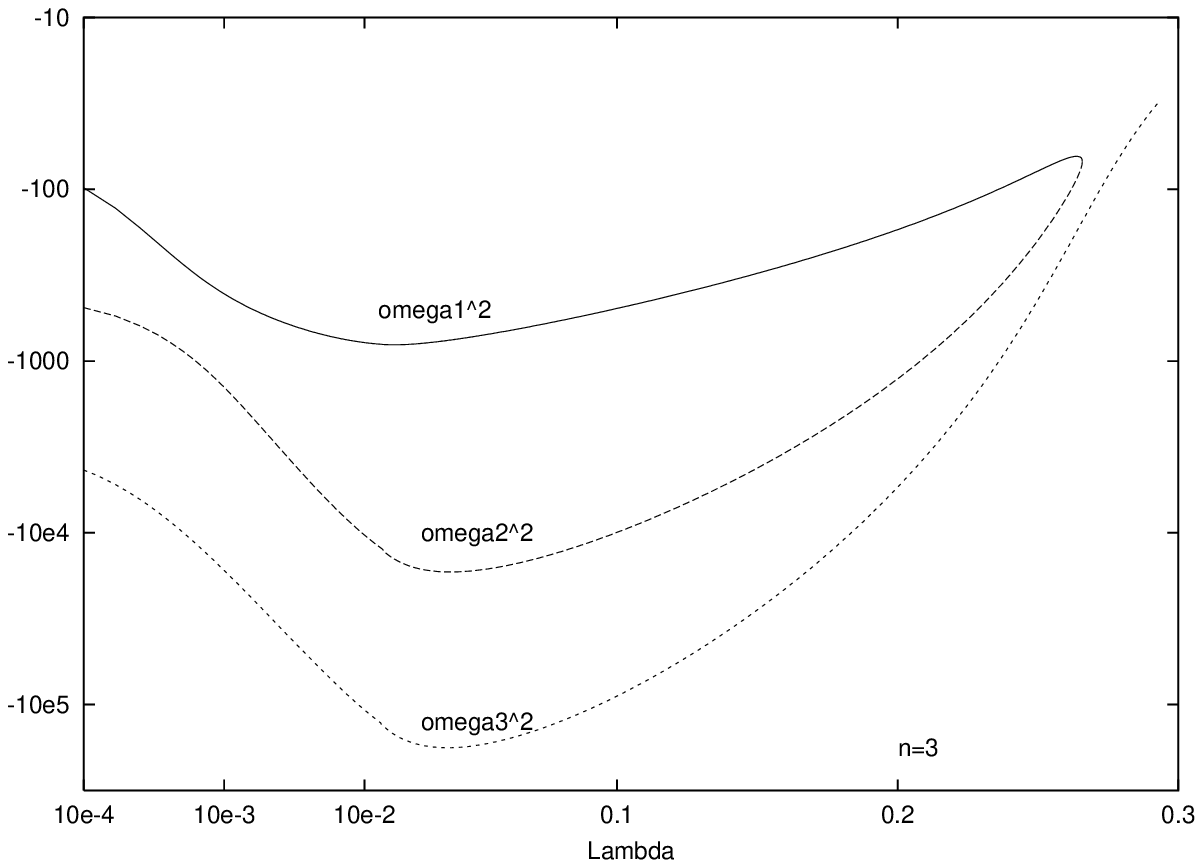}}%
\hspace{5mm}%
\hss} }

\caption{The eigenvalues of the unstable modes plotted as functions of $\Lambda$ for the
$n=1,2,3$ solutions.}
\end{figure}
\end{document}